\documentclass[aps, twocolumn,showpacs,preprintnumbers,amsmath,amssymb]{revtex4}

\usepackage{graphicx}
\usepackage{dcolumn}
\usepackage{bm}
\usepackage{epstopdf}
\usepackage{color}

\begin{document}
\preprint{APS/123-QED}

\title{Modification of the $^3$He Phase Diagram by Anisotropic Disorder}

\author{R.G. Bennett$^{\dagger a}$, N. Zhelev$^{\dagger a}$, E.N. Smith$^{\dagger}$,
J. Pollanen$^{\ddag}$, W.P. Halperin $^{\ddag}$, and J.M. Parpia$^{\dagger}$}

\affiliation{$^{\dagger}$Department of Physics, Cornell University, Ithaca, NY, 14853 USA}
\affiliation{$^{\ddag}$Department of Physics and Astronomy, Northwestern University, Evanston, IL 60208, USA}

\date{\today}

\begin{abstract}

 Motivated by the recent prediction that uniaxially compressed aerogel can stabilize the anisotropic A phase over the isotropic B phase, we measure the pressure dependent superfluid fraction of $^3$He entrained in 10\% axially compressed, 98\% porous aerogel.  We observe that a broad region of the temperature-pressure phase diagram is occupied by the metastable A phase. The reappearance of the A phase on warming from the B phase, before superfluidity is extinguished at $T_c$, is in contrast to its absence in uncompressed aerogel. The phase diagram is modified from that of pure $^{3}$He, with the disappearance of the polycritical point (PCP) and the appearance of a region of A phase extending below the  PCP of bulk $^{3}$He, even in zero applied magnetic field. The  expected alignment of the A phase texture by compression is not observed.
\end{abstract}

\pacs{61.43.Fs, 62.65.+k, 63.50.-x, 62.25.Fg}
\maketitle

The introduction of controlled disorder into an otherwise pristine material enables the investigation of that material's susceptibility to modification of its properties. Silica aerogel, a dilute self supporting structure, is the only means of introducing an impurity into a three dimensional quantum fluid\cite{porto,halperin}. By introducing disorder on a scale of order the zero temperature coherence length, the pairing is disrupted leading to a suppression of the order parameter. This has several consequences including a significant modification of the phase diagram where a quantum phase transition is known to occur\cite{matsumoto}, gapless superfluidity\cite{sauls} (manifested by finite excitation densities even at $T=0$ in the heat capacity\cite{halperin2}, thermal conductivity\cite{lancaster} and superfluid density\cite{porto}), as well as an alteration of the delicate balance between competing phases\cite{osheroff1,gervais,osheroff2,osheroff3,nazaretski,Lee2}. Furthermore, deliberate anisotropy can be introduced into the aerogel by compressing it\cite{Aoyama, Fomin}, thus altering the correlation length along a particular axis. Such a compressed aerogel can be optically characterized before and after compression\cite{Pollanen1}.

The fact that $^3$He has a $p$-wave paired state, can provide insight into the behavior of superconducting systems with complex order parameters. However, the simplicity of the Fermi surface of $^3$He ensures that the quantum fluid is only a starting point for understanding the behavior of its sister electronic systems\cite{rice}. Multicomponent systems such as UPt$_3$\cite{joynt,adenwalla} have been shown to display internal phase transitions, and moreover it is known that sample purity and quality affect the onset of superconductivity. Triplet superconducting states are in a special class; various of these states  are chiral, spontaneously breaking time reversal symmetry in zero magnetic field.  One of these is the A phase of superfluid $^3$He, and others have been proposed for UPt$_3$ and for Sr$_2$RuO$_4$\cite{rice2}. In bulk superfluid $^3$He, the A-phase is a chiral phase, stabilized by strong coupling effects.  At issue is the question, raised theoretically in \cite{thuneberg1,Aoyama} whether anisotropic scattering from aerogel impurities can also be a mechanism for stabilizing anisotropic states such as the A-phase, given that impurity scattering decreases the effects of strong coupling \cite{review}. These effects of impurities are thus relevant to our understanding of the states proposed for chiral superconductors.  It is possible that the known anisotropic scattering in UPt$_3$ might be a case in point \cite{Kycia}. Impurity studies with anisotropic scattering are therefore of importance in unconventional pairing systems, and insight into phase stability  of these more complex systems can be gained through investigation of quenched anisotropic disorder in $^3$He.

Experiments on $^3$He in uncompressed aerogel reveal that the A phase is reliably nucleated upon cooling from the normal state\cite{osheroff1,gervais,osheroff2,nazaretski,Lee2} and persists over a wide range in temperature and pressure. The A to B transition has finite width, typically ~30 $\mu$K, with the A-B interface most likely pinned by the presence of the aerogel \cite{osheroff2,nazaretski}. Upon warming, the B phase is observed up to the superfluid transition, $T_{ca}$, with the width of the B to A transition spanning the width of the superfluid transition itself. Reports of the A phase reappearing below $T_{ca}$ \cite{Lee2} are limited to the region very close to $T_{ca}$ {\it e.g.}\cite{osheroff1}, and no systematic pressure dependence has been reported. Furthermore, there have been proposals that the disorder inherent in uncompressed aerogel might favor the Larkin-Imry-Ma state over the A phase due to local variations in the order parameter \cite{volovik,volovik2}. More recently Dmitriev and Volovik have demonstrated remarkable control over the A phase and variants using compressed and stretched aerogels; experiments were carried out in the metastable or supercooled A phase but there has been no observed alteration of the morphology of the phase diagram of $^3$He in compressed {\it vs.} uncompressed aerogel\cite{davis,dmitriev1,bunkov1,bunkov2}.

\begin{figure}
\begin{center}
\includegraphics[%
width=1\linewidth,
keepaspectratio]{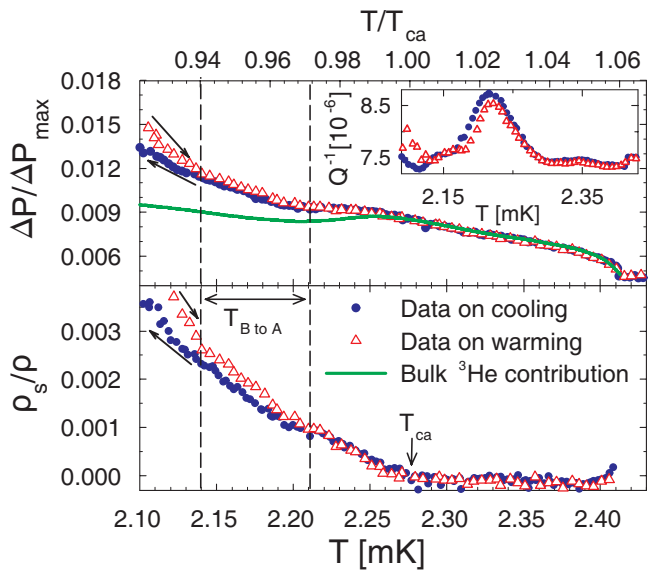}
\end{center}
\caption{(Color online). (Upper panel) Period shift $vs$ temperature near the superfluid transition at 31.9 bar showing data taken while cooling (filled (blue) circles) and warming (open (red) triangles), the latter obtained after cooling into the B phase. The solid (green) line is a fit of the bulk superfluid period shift. Inset shows the corresponding dissipation ($Q^{-1}$) and broad 4$^{\rm th}$ sound resonance whose hysteresis is likely due to bulk A phase textural effects.
(Lower panel) After subtraction of the bulk superfluid contribution we show the superfluid density of $^3$He in 98\% open aerogel under 10\% axial compression. The arrow designates the onset of superfluidity and dashed lines define the width of the $B\rightarrow A$ transition.}
\label{Fig1}
\end{figure}

 In this letter we report results using a torsion oscillator to explore the phase diagram\cite{AB} without the application of a magnetic field \cite{dmitriev1,bunkov1,bunkov2} with a high precision volume rather than surface sensitive method\cite{davis}, thus extending previous work. The 98\% open aerogel was grown in a stainless steel shell and then compressed by 10\% to a height of 400 $\mu$m.  Before and after compression the aerogel sample was characterized by optical birefringence\cite{Pollanen1}. The shell was then mounted in an epoxy head where it served as the inertial mass of a double torsion pendulum, with the axis of compression along the torsion rod. We note the presence of bulk fluid in two regions around the shell, which we modeled as channels of height 30 $\mu$m and 400 $\mu$m contributing 3.2\% and 0.8\% respectively to the moment of inertia, parameters that were determined from measurements in the normal state\cite{BennettJLTP}. Subtraction of the $T=0$ empty cell period from the period of the filled cell extrapolated to its ``fully locked" value allows us to determine the $^3$He contribution to the period shift, $\Delta P_{\rm max}$. We measured the temperature dependent background (period ($P$) and dissipation ($Q^{-1})$) of the empty oscillator, while maintaining a small constant amplitude ($\approx$0.1 nm) to avoid non-linear behavior of the torsion rod. Thermometry was provided by a $^3$He melting curve thermometer external to the cell \cite{PLTS} and a quartz tuning fork 
immersed in the  $^3$He that provided signatures of bulk $T_c$ and A-B transitions and verified negligible thermal gradients between the experiment and thermometers.

$P(T)$ and $Q^{-1}(T)$ were monitored while cooling and warming through the various transitions. Data for the period shift ($\Delta P=P(T)-P(T_c)$) obtained at 31.9 bar, and close to $T_c$, is shown in Fig.~1, together with a fit for $\Delta P$ expected for the bulk fluid, using published results for $^3$He viscosity and superfluid density\cite{eandpPRL,parpiajltp
}. We also account for the non-monotonic period shift from bulk superfluid 4$^{\rm th}$ sound resonances that cross the oscillator frequency (inset to Fig.~1 and supplementary material). The data depart from the bulk behavior below 2.28 mK, marking the onset of superfluidity of the $^3$He in aerogel. At a lower temperature, 2.21 mK, we note that the warming and cooling data converge, marking the location of the completion (on warming) of the B $\rightarrow$ A transition of $^3$He in anisotropic aerogel.

The superfluid fraction of the $^3$He in the aerogel ($\rho_s/\rho$) is given by $(P(T)-P(T_{ca}))/(\Delta P_{\rm max})$ after subtracting the bulk superfluid contribution (solid (green) line in Fig.~1).
The superfluid fraction near $T_{ca}$ is shown in Fig.~1 and in Fig.~2 for 31.9, 25.7, 21.9 and 15.2 bar.

\begin{figure}
\begin{center}
\includegraphics[%
 width=0.95\linewidth,
  keepaspectratio]{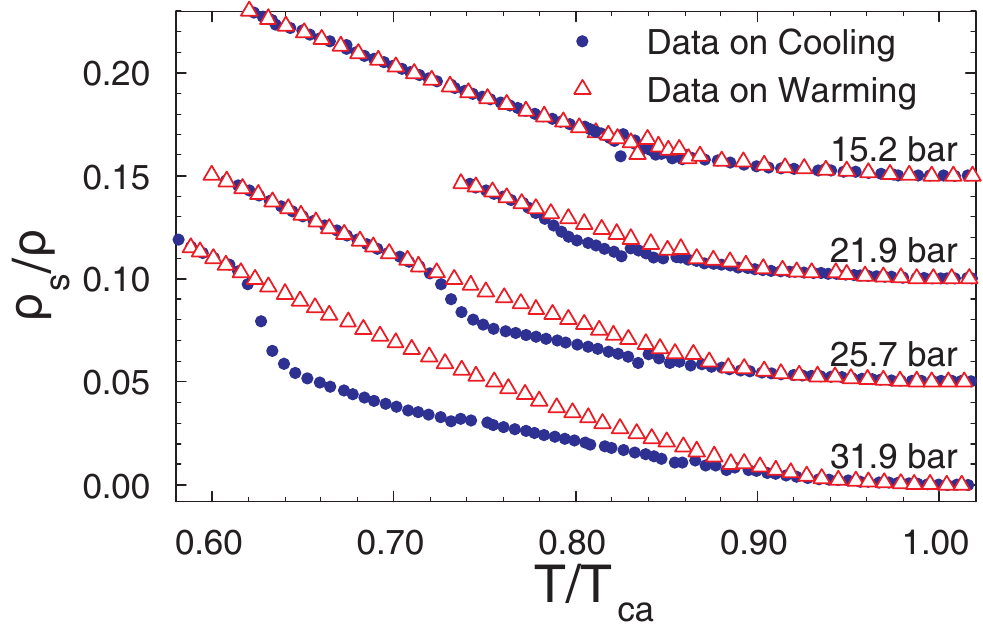}
\end{center}
\caption{(Color online). The superfluid fraction ($\rho_s/\rho$) $vs$ temperature after the bulk superfluid contribution is subtracted. The (blue) circles and (red) triangles represent data obtained while cooling and warming respectively at various pressures (offset by 0.05 for clarity). Resonances where the slow mode (a composite fourth sound-like mode\cite{golov}) and torsional oscillator frequency cross are visible near 0.84 $T_{ca}$.  The metastable region occupied by the A phase on cooling is emphasized; the conversion from A to B phase in compressed aerogel occurs over a band $\sim70$ $\mu$K wide
and shows that $\rho_{s}^A< \rho_{s}^B$.}
\label{Fig2}
\end{figure}

It is well established that in bulk $^3$He, $\rho_s^A/\rho$ is a tensor quantity, with $\rho_{s\bot}^A>\rho_{s}^B>\rho_{s\|}^A$\cite{giannetta}. Because the compression of the aerogel is aligned with the torsion axis, the A phase order parameter should be aligned with its nodes oriented along the same axis; thus the oscillator should sample the $\rho_{s\bot}^A$ component of the superfluid tensor and one would expect to see a reduction in $\rho_s/\rho$ when entering the B phase from the A phase. In disordered $^3$He, flow alignment of $\rho_{s}^A$ has been demonstrated by the Lancaster group \cite{Fisher}, who drove a composite aerogel-wire resonator to large amplitude, aligning the $l$ texture along the flow direction; the alignment persisted even when the amplitude was reduced. In this work we observe $\rho_s^B/\rho_s^A>1$ (nearly identical to that seen earlier in uncompressed aerogel\cite{nazaretski}). We conclude that the expected alignment of $l$ in the A phase by compression\cite{volovik2} is $not$ observed. We attempted flow alignment in the A phase but the maximum achievable velocity was $\approx30$ times smaller than that shown to be necessary in Ref.~\cite{Fisher} and we observed no change in the width or $\rho_s$ of the metastable A phase.

\begin{figure}
    \includegraphics [%
  width=0.96\linewidth,
  keepaspectratio]{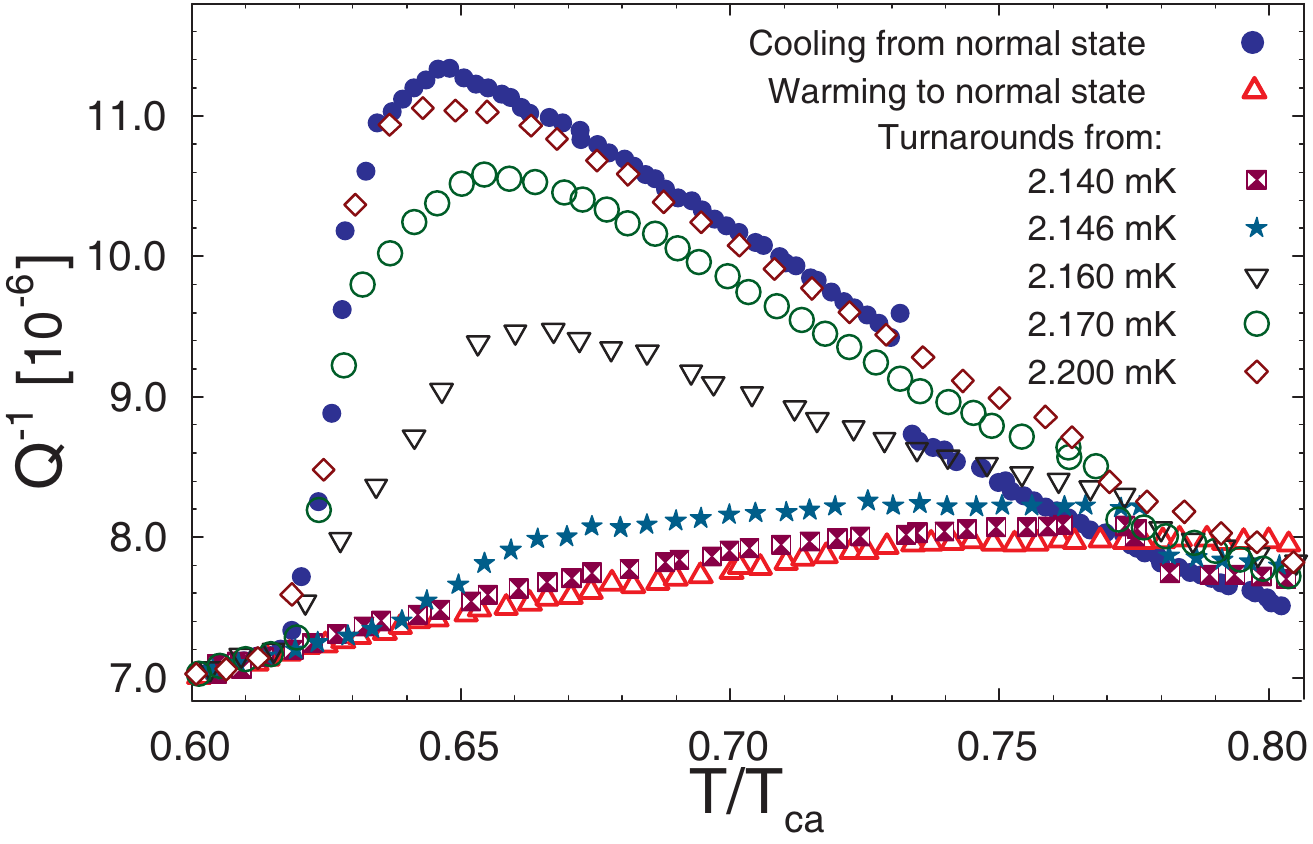}
    \caption{(Color online). Dissipation data for turn around measurements at 31.9 bar, in the vicinity of the $A\rightarrow B$ transition. The (blue) circles and (red) triangles represent data taken on cooling from, and warming to, $T_{ca}$ respectively. The intermediate points represent data taken on cooling, following turn arounds at the temperatures given in the legend. The abrupt jump near 0.73$T/T_{ca}$  in solid circles is the signature of the bulk $A\rightarrow B$ transition, also seen in the fork thermometer.}
    \label{figure3}
    \end{figure}

In the bulk, the A phase exhibits a number of interesting behaviors. It is highly metastable and supercools in the presence of clean surfaces down to at least 0.15$T_c$ at high pressure\cite{schiffer}, provided that extrinsic nucleation centers can be reduced. It is likely that the interfacial surface energy prevents nucleation of the B phase, but once the B phase is nucleated by extrinsic mechanisms ({\it e.g.} \cite{bart}), the transition proceeds rapidly to completion. Upon warming, the bulk displays no superheating, presumably because ``seeds" of the A phase are present and there is no barrier to nucleation. Thus the extent of the ``equilibrium" bulk A phase is defined by the B$\rightarrow$A transition temperature (dotted line in Fig.~4). In contrast, in the so-called ``dirty" or impurity dominated $^3$He, the $A\rightarrow B$ and $B\rightarrow A$ transitions have finite width, which has been attributed to pinning and possible inhomogeneities\cite{nazaretski}.

\begin{figure}
    \includegraphics [%
  width=1.0\linewidth,
  keepaspectratio]{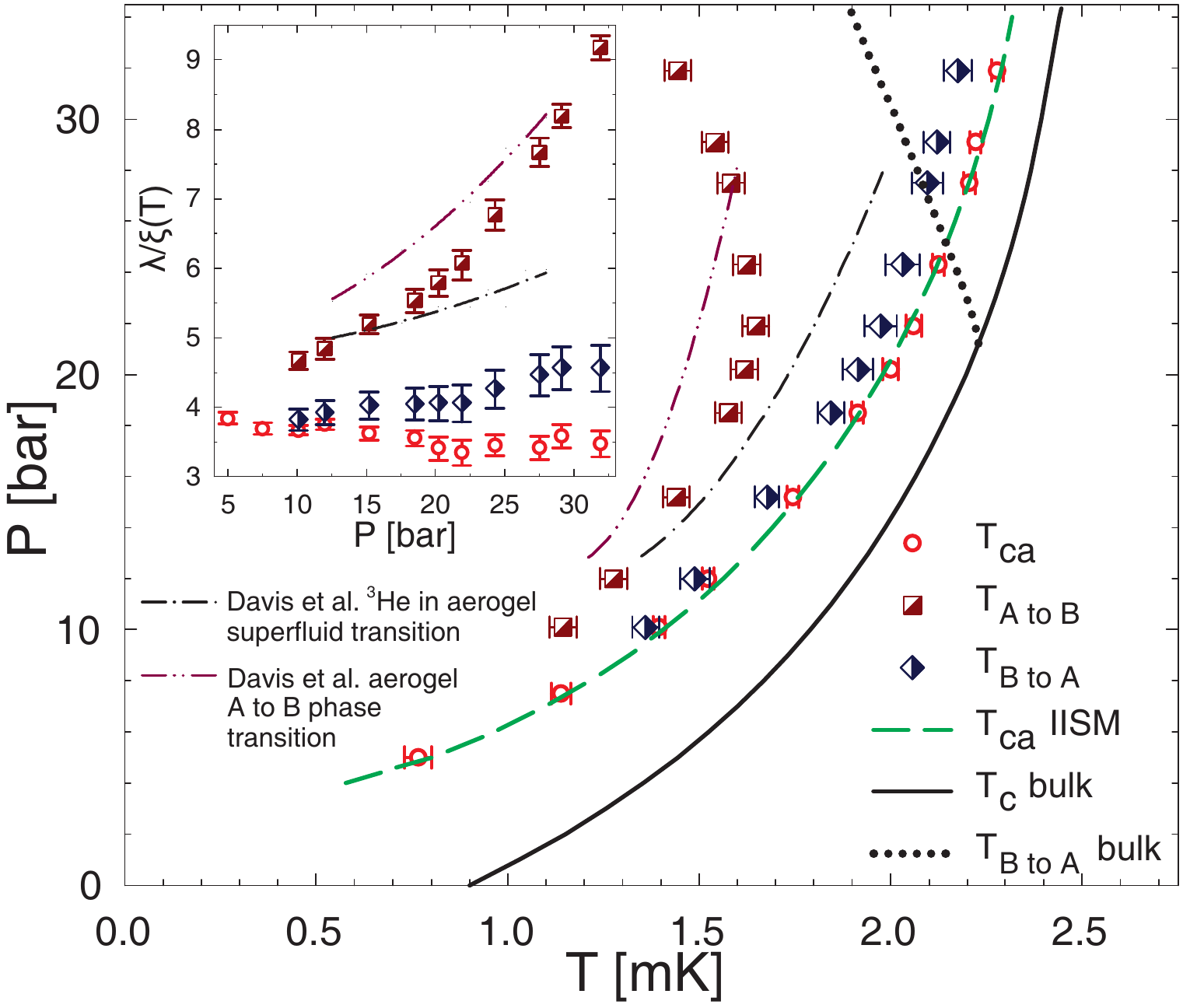}
   \caption{(Color online). The phase diagram for superfluid $^3$He in 10\% axially compressed, 98\% open aerogel. The suppressed onset of superfluidity is marked by (red) circles, and the low temperature boundary of the metastable A phase marked by (brown) squares. The $B\rightarrow A$ transition is denoted by (blue) diamonds. The solid line represents $T_c$ and the dotted line represents $T_{AB}$ for bulk superfluid $^3$He. The polycritical point (junction of the dotted and solid lines), where the bulk A, B and normal states coexist, is removed by anisotropic disorder. The dashed line represents $T_{ca}$ as calculated using the IISM model\cite{SS}. We also show fits through the data of Davis {\it et al.}\cite{davis} who did not observe a distinct $B\rightarrow A$ transition. Inset: the phase diagram for both data sets as a plot of the ratio of the inferred mean free path, $\lambda/\xi$(T), against the pressure. }
    \label{figure4}
    \end{figure}

The observed width of the $A\rightarrow B$ transition suggests \cite{nazaretski} that the $B\rightarrow A$ transition might also be wide. Thus we sought to resolve the conversion of B phase to A phase by conducting a series of ``turn around'' measurements warming the cell
at 30-60 $\mu$K/hr followed by a period of several hours where 
the cell warmed slowly (2-3 $\mu$K/hr) 
until we reached our target temperature.  The cell was then cooled again at 30-60 $\mu$K/hr back into the B phase. 
It was evident that the $Q^{-1}$ signature at the $A\rightarrow B$ transition was a very sensitive indicator of the presence of A phase, B phase or an admixture of the two\cite{AB}. Data for $Q^{-1}(T)$ at 31.9 bar, as turn arounds proceeded to successively higher temperatures, are shown in Fig.~3. In this sample the width of the $B\rightarrow A$ transition (vertical dashed lines in Fig.~1) can be seen to be $\approx70$ $\mu$K. The nucleation of the B phase in aerogel is not influenced by the adjacent bulk B phase\cite{gervais} evidenced by the persistence of A phase below the bulk $A\rightarrow B$ transition (Fig. 3). We cannot infer the distribution of the A and B phases.

Similar measurements were carried out at several pressures extending down to the saturated vapor pressure. Hysteresis in the superfluid density on warming and cooling was observed at pressures down to 10 bar with a distinct separation of the $B\rightarrow A$ transition and $T_{ca}$ discernable down to 10 bar. Thermometry precluded a definitive observation of superfluidity at 2.6 bar, with no superfluidity observed down to 0.5 mK at this pressure. Similarly, at the saturated vapor pressure no superfluidity was discerned. In general, our data shows that the 98\% aerogels grown by the Northwestern group suppress the superfluidity of the dirty $^3$He by a smaller amount than those fabricated by Mulders\cite{nazaretski}. Similar conclusions were drawn in the work of Davis {\it et al.}\cite{davis}.

The presence of the semi-rigid aerogel introduces elastic scattering sites that limit the quasiparticle mean free path to a pressure independent length $\lambda$. Various models have been invoked to adapt the physics of the suppression of superconductivity by magnetic impurities, first calculated by Abrikosov and Gorkov\cite{AG}, to ``dirty" $^3$He, starting with the work by Thuneberg {\it et al.}\cite{thuneberg1}. The  most approachable and successful  way of describing the suppression of $T_c$ by aerogel is given by the inhomogeneous isotropic scattering model (IISM). Sauls and Sharma have presented a readily applicable phenomenological model \cite{review,SS} which introduces a dimensionless scaling parameter $\zeta_a$=$\xi_a/\lambda$, where $\xi_a$ is the aerogel particle-particle correlation length. A second important length scale is provided by the pressure dependent zero temperature coherence length $\xi_0=\hbar v_F/(2\pi k_BT_c)$ that is used to define a pairbreaking parameter $\hat{x}=\xi_0/\lambda$. By fitting to the data we determine $\lambda=155$ nm and $\xi_a=85$ nm.  The fit is shown in Fig.~4. We also show the earlier measurements of Davis {\it et al.}\cite{davis} that did not observe the reappearance of the A phase on warming, for which $\lambda=140$ nm and $\xi_a=85$ nm. We note that acoustic impedance measured in transverse sound is sensitive to the surface Andreev bound states of the superfluid while the torsional oscillator probes the entire sample, accounting for the differences between the two experiments. Although a correct model should allow for anisotropic scattering it is reasonable to assume that $\lambda$ and $\xi_a$, obtained from the IISM are directionally averaged values.

We use $\lambda$ to scale the temperature dependent coherence length, $\xi(T)=[7\zeta(3)/48]^{1/2}(\hbar v_F/\pi k_BT_c)(1-T/T_c)^{-1/2}$. This approximately linearizes the suppression of $T_{ca}$ (inset to Fig. 4). The reappearance of the A phase on warming is weakly pressure dependent and the supercooling of the A phase is more strongly so. Fig. 4 conclusively shows the stabilization of the A phase by anisotropic disorder and suppression of the polycritical point.

In conclusion, in this letter we show that the addition of anisotropic disorder in the form of a uniaxially compressed aerogel, has the effect of enhancing the width of the metastable A phase, and increasing the width of the region where the A phase is the lowest energy state (even when compared to an identical uncompressed sample\cite{pollanennew}), all in zero magnetic field. The reappearance of the A phase from the B phase on warming is manifested at pressures well below the bulk polycritical point. Due to the supercooling of the A phase and possible superheating or pinning (as also seen in the bulk\cite{bart}) of the B phase,  the ``equilibrium" A-B phase boundary cannot be identified by zero field measurements alone \cite{gervais,moon}.

In this experiment, we have demonstrated that the phase diagram is significantly altered, with the removal of the polycritical point. The expected alignment of the angular momentum $l$ by compression is not observed. The experimental finding that anisotropic disorder affects the stability of competing phases, as well as suppressing the superfluidity, may well be applicable in understanding the evolution of phase diagrams in exotically paired superconducting systems.

We acknowledge support from the  NSF under DMR-0806629 at Cornell and DMR-1103625 at Northwestern. $^a$ R.G. Bennett and N. Zhelev contributed equally.


\begin{thebibliography}{999}
\bibitem{AB}
Torsion oscillator measurements map phase transitions as changes in the frequency and $Q^{-1}$ but cannot identify specific phases.  We follow generally accepted identification of the A and B phases \cite{osheroff1} and their coexistence \cite{osheroff2}.
\end{thebibliography}
\end{document}